\documentclass[
reprint,
superscriptaddress,
 amsmath,amssymb,
 aps,
floatfix
]{revtex4-1}

\usepackage{comment}
\usepackage{verbatim}
\usepackage{graphicx}
\usepackage{dcolumn}
\usepackage{bm}
\usepackage{hyperref}
\usepackage{inputenc}
\usepackage{amsfonts}
\usepackage{multirow}

\newcommand{\SM}{Supplemental Material~\cite{SM}}

\begin{document}

\preprint{APS/123-QED}

\title{Systematic comparison of graph embedding methods in practical tasks}

\author{Yi-Jiao Zhang}
\affiliation{Institute of Computational Physics and Complex Systems, Lanzhou University, Lanzhou, Gansu 730000, China}

\author{Kai-Cheng Yang}
\affiliation{Center for Complex Networks and Systems Research, Luddy School of Informatics, Computing, and Engineering, Indiana University, Bloomington, Indiana 47408, USA}

\author{Filippo Radicchi}
\affiliation{Center for Complex Networks and Systems Research, Luddy School of Informatics, Computing, and Engineering, Indiana University, Bloomington, Indiana 47408, USA}
\email{filiradi@indiana.edu}

\date{\today}

\begin{abstract}
Network embedding techniques aim at representing structural properties of graphs in geometric space.
Those representations are considered useful in downstream tasks such as link prediction and clustering. 
However, the number of graph embedding methods available on the market is large, and practitioners face the non-trivial choice of selecting the proper approach for a given application.
The present work attempts to close this gap of knowledge through a systematic comparison of eleven different methods for graph embedding. 
We consider methods for embedding networks in the hyperbolic and Euclidean metric spaces, as well as non-metric community-based embedding methods.
We apply these methods to embed more than one hundred real-world and synthetic networks. Three common downstream tasks --- mapping accuracy, greedy routing, and link prediction --- are considered to evaluate the quality of the various embedding methods.
Our results show that some Euclidean embedding methods excel in greedy routing.
As for link prediction, community-based and hyperbolic embedding methods yield overall performance superior than that of Euclidean-space-based approaches.
We compare the running time for different methods and further analyze the impact of different network characteristics such as degree distribution, modularity, and clustering coefficients on the quality of the different embedding methods.
We release our evaluation framework to provide a standardized benchmark for arbitrary embedding methods.
\end{abstract}

\maketitle

\section{Introduction}

Representing complex networks in latent space, or network embedding, has generated a growing interest from multiple disciplines~\cite{boguna2021network,hamilton2017representation,goyal2018graph}. 
From a theoretical point of view, the geometric representation of a network may provide an intuitive explanation of key properties of real-world systems such as structural features~\cite{papadopoulos2012popularity}, navigability~\cite{boguna2009navigability, gulyas2015navigable}, and robustness~\cite{kleineberg2017geometric, osat2020kcore}; when it comes to applications, network embedding can be useful for graph analysis tasks like visualization~\cite{van2008visualizing}, link prediction~\cite{liben2007link}, and graph clustering~\cite{ding2001min, tandon2021community}. 

Many embedding methods use Euclidean space as their target space. Euclidean embedding is intuitive and can immediately be used in standard machine learning algorithms~\cite{hamilton2017representation,goyal2018graph}. 
However, network embedding methods are not limited to Euclidean space. 
For example, many recent approaches represent networks in hyperbolic space, where properties like hierarchy and heterogeneity can be easily captured~\cite{papadopoulos2015network_2,garcia2019mercator,klimovskaia2020poincare,keller2020hydra,kitsak2020link}.
Community structure can be seen as an alternative approach to network embedding
in non-metric spaces~\cite{faqeeh2018characterizing}. 

The existence of so many available and diverse embedding techniques presents a challenge for practitioners when they have to choose the proper method for the application at hand. Standardized tests for systematic comparison among methods are lacking. The effectiveness of embedding methods is generally measured on limited types of tasks and small corpora of real-world networks. As a result, gauging the relative performance of a method with respect to another is difficult.

In this work, we address this gap of knowledge by performing a systematic comparison of representative embedding methods. 
We consider five hyperbolic embedding methods (HyperMap~\cite{papadopoulos2015network_1, papadopoulos2015network_2}, Mercator~\cite{garcia2019mercator}, Poincar\'e maps~\cite{klimovskaia2020poincare}, Hydra~\cite{keller2020hydra}, and HyperLink~\cite{kitsak2020link}), four Euclidean-space-based approaches (Node2vec~\cite{grover2016node2vec}, Laplacian Eigenmap~\cite{belkin2001laplacian}, HOPE~\cite{ou2016asymmetric}, and Isomap~\cite{tenenbaum2000global}), and the two variants (relying on  Louvain~\cite{blondel2008fast} and Infomap~\cite{rosvall2008maps}) of the non-metric community embedding method~\cite{faqeeh2018characterizing}.
We apply these methods to embed more than one hundred real-world and synthetic networks. 
Three downstream tasks, i.e., mapping accuracy, greedy routing, and link prediction, are considered to evaluate the quality of the various embedding methods. 
We assess how the performance of the various methods is affected by network characteristics such as degree distribution, modularity, and average clustering coefficient.
The various methods are also compared in terms of their computational complexity and their number of tunable parameters. 

Our findings indicate that Euclidean embedding methods such as Node2vec and Isomap represent the overall best choice for practitioners as they yield decent performance in all tasks.
Hyperbolic embedding methods excel in link prediction; however, their high computational complexity impedes their application to large-scale networks.
Community-based methods behave similarly to hyperbolic embedding methods, but they require 
a lower computational demand.
Our systematic analysis includes many different embedding methods. However for obvious reasons, we could not include all methods that are currently available on the market or that will be developed in the future. To ease the analysis of arbitrary embedding methods under our proposed experimental setting, we made it publicly available at \url{https://github.com/yijiaozhang/hypercompare}.

\section{Graph visualization}

\begin{figure*}[!htb]
\includegraphics[width=1\textwidth]{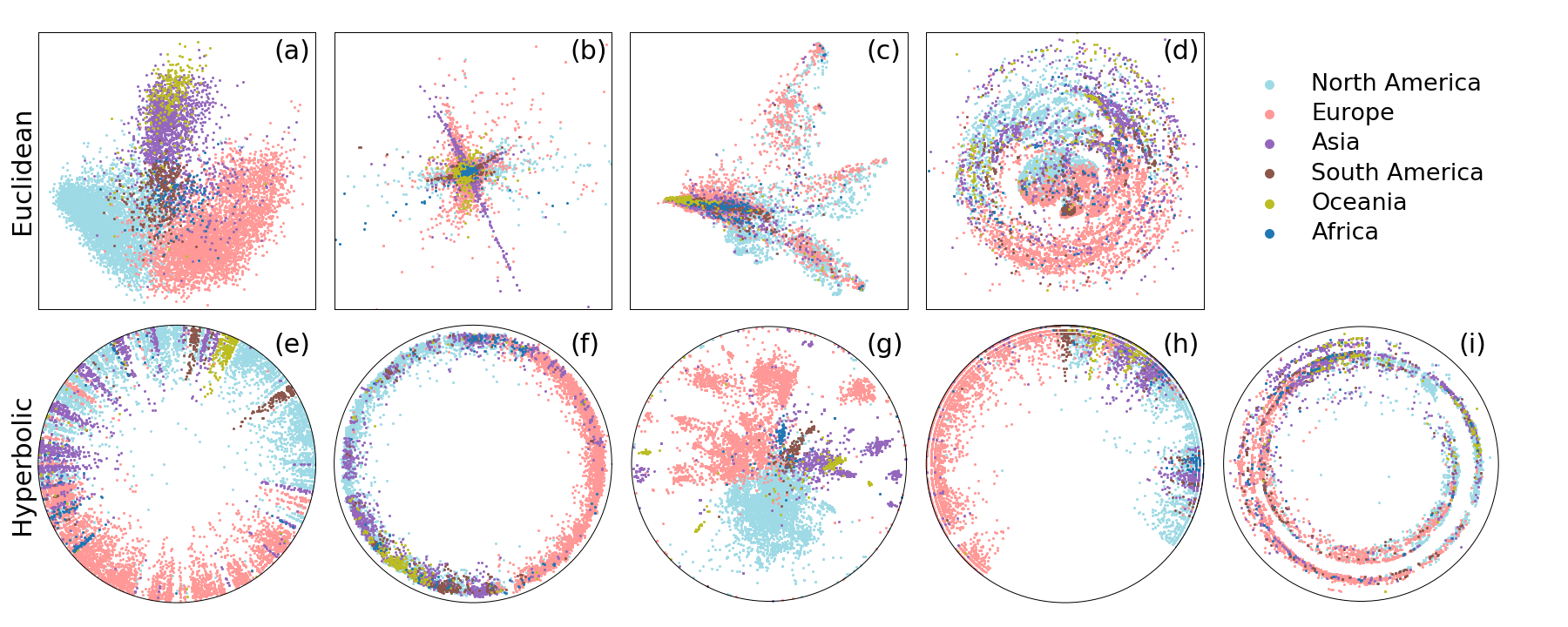}
\caption{\label{fig:illustration} 
\textbf{Geometric embedding of the Internet.}
We display the visualization of the autonomous system (AS) Internet network in Euclidean space inferred by (a) Node2vec, (b) HOPE, (c) LE, (d) Isomap, and in the Euclidean projection of the hyperbolic embedding as inferred by (e) HyperMap, (f) Mercator, (g) HyperLink, (h) Poincar\'e maps, (i) Hydra. The color of a point is representative for the continent where the respective AS is located in. For clarity of the visualization, only nodes with degree larger than one are shown. For the visualization of Node2vec, HOPE, and LE, we first get the coordinates with dimension $d = 128$, and then use PCA to obtain a two-dimensional projection. For the other methods, we directly plot their two-dimensional embeddings.}
\end{figure*}

To qualitatively illustrate differences between different network embedding methods,
we display graphical visualizations 
produced by the various methods for the same network topology,  i.e., the autonomous system (AS) Internet network~\cite{boguna2010sustaining}. The network contains $N = 23,748$ nodes and $E = 58,414$ edges. Visualizations are displayed in Fig.~\ref{fig:illustration}. 

It is important to stress that all visualizations are performed in the two-dimensional Euclidean space, thus the original embedding is projected in this space using some ad-hoc recipes.
For example, to yield decent embedding results, a high embedding dimension is required for Node2vec, LE, and HOPE.
We therefore first learn their 128-dimensional embeddings and then use principal component analysis (PCA) to project the results into the two-dimensional plane of the figure.
The visualization by Isomap is obtained directly by setting the embedding dimension to two.
For hyperbolic embedding methods, we represent the embedded nodes with their polar coordinates or Poincar\'e coordinates and plot them in the two-dimensional Euclidean projection of the Poincar\'e disks.
Finally, despite their potential use in graph drawing, we exclude the non-metric community-based embedding methods from the qualitative analysis in order to avoid  the use of sophisticated projections in the two-dimensional Euclidean space.

To help the readers making sense of the visualizations, we color the autonomous systems, i.e., the nodes of the network, according to the continent where they are located in.
We can see that, although different embedding methods yield drastically different visualizations, all of them can preserve geographic proximity to some extent, i.e., nodes within the same continent are often close one to the other in the visualizations.
If we consider polar coordinates for all the embeddings (using the geometric center as the origin for Euclidean embeddings), it becomes clear that the angular coordinates encode the community structure of the graph~\cite{faqeeh2018characterizing,zhang2021model}.
The radial coordinates, on the other hand, often convey centrality information.
For example, hyperbolic embedding methods like HyperMap, Mercator, and HyperLink use the radial coordinates to represent the degree of the nodes~\cite{papadopoulos2015network_2,garcia2019mercator,kitsak2020link}.
The distance from the geometric center of a node in the Isomap, Hydra, and Poincar\'e maps embedding is correlated with its closeness centrality~\cite{zhang2021model}. For other methods, such association is still unclear. We find that the distance from the geometric center of a node obtained by LE and HOPE is highly correlated with its closeness and eigenvector centrality, respectively.
Detailed analysis and discussion can be found in the \SM.

\section{Performance in downstream tasks}

We now use downstream tasks to quantify the embedding quality of different methods. 
Specifically, we measure their performance in mapping accuracy, greedy routing, and link prediction.
These tasks are conducted on 72 real-word networks representing social, biological, technological, transportation, and communication systems. Details of these networks are included in the \SM.

To summarize the results from all the networks for an embedding method on a task, we produce the complementary cumulative distribution function (CCDF) of a performance metric and calculate the area under the CCDF curve (CCDF-AUC) as the overall score.
The CCDF-AUC matches the average value of the performance metric over the entire corpus of real-world networks and higher CCDF-AUC values indicate better overall performance.

Some embedding methods have free parameters that could affect the measured value of the performance metric.
We tune the parameters for each method to find the optimal value of the overall performance, and use these  parameter values for all networks, see Sec.~\ref{sec:emb_methods} for details.

\begingroup
\squeezetable
\begin{table*}[!htb]
\caption{
{\bf Network embedding methods.} 
We summarize some features of the embedding methods considered in this paper and the key results.
From left to right, we report: name of the method, the target embedding space (space), programming language of the publicly available implementation (lang.), network structural information preserved by the method (struct. preserv.), computational complexity (complexity), CCDF-AUC for mapping accuracy (mapp. acc.), CCDF-AUC for greedy routing (greedy rout.), and CCDF-AUC for link prediction (link pred.).
For each task, we highlight in bold face the CCDF-AUC values of the top three embedding methods.
In the expressions of the computational complexity, $N$ is the number of the nodes, $E$ is the number of the edges, $d$ is the embedding dimension, $C$ is the cost to compute each entry of the shortest path length matrix, $e$ is the number of epochs (we set $e = 1,000$), 
$b = \min \{512, N/10\}$
is the batch size, $m$ is the number of node layers, and $\langle k \rangle$ is the average degree of the network.
More details about the methods can be found in Sec.~\ref{sec:emb_methods}.
The CCDF-AUC values are generated by aggregating the performance on 72 real-world networks for mapping accuracy and greedy routing.
For link prediction, the CCDF-AUC values are computed on a subset of 46 real-world networks with size larger than 300.
The CCDF-AUC values for HyperLink are marked with * because the method is unable to embed several networks.
Restricting the analysis on the subset of real-word networks that HyperLink can process yields qualitatively similar results in all three tasks (see \SM for details).
}
\label{tab:embedding_methods_summarize}
\begin{ruledtabular}
\begin{tabular}{r|llll|ccc}
\textrm{Method} & \textrm{Space} & \textrm{Lang.} & \textrm{Struct. preserv.} & \textrm{Complexity} & \textrm{Mapp. acc.} & \textrm{Greedy rout.} & \textrm{Link pred.} \\
\hline
Node2vec \cite{grover2016node2vec} & Euclidean & Python & Tunable & $O(dN)$ &  0.561 & \textbf{0.818} & 0.787 \\
HOPE \cite{ou2016asymmetric} & Euclidean & Python & Global & $O(d^2 E)$ & 0.575 & \textbf{0.703} & 0.769 \\
Laplacian Eigenmap (LE) \cite{belkin2001laplacian} & Euclidean & Python & Local & $O(d^2 E)$ &  0.464 & 0.566 & 0.749 \\
Isomap \cite{tenenbaum2000global} & Euclidean & Python & Global & $O(CN^2 + d N^2)$ &  \textbf{0.858} & \textbf{0.861} & 0.848 \\
\hline
HyperMap \cite{papadopoulos2015network_1} & hyperbolic & C++ & Local  & $O(N^2)$ &  0.388 & 0.584 & 0.840 \\
Mercator \cite{garcia2019mercator} & hyperbolic & Python & Local  & $O(N^2)$ & 0.557 & 0.530 & \textbf{0.902} \\
HyperLink \cite{kitsak2020link} & hyperbolic & C++ & Local & $O(m \langle k \rangle N^2)$ & 0.526* & 0.596* & 0.891* \\
Poincar\'e maps \cite{klimovskaia2020poincare} & hyperbolic & Python & Global  & $O(N^2 + ebN)$ & \textbf{0.628} & 0.494 & 0.822 \\
Hydra \cite{keller2020hydra} & hyperbolic  & R & Global & $O(N^{\alpha}),  \alpha > 2$ & \textbf{0.799} & 0.683 & 0.846 \\
\hline
Community embedding (Infomap) \cite{faqeeh2018characterizing} & non-metric & Python & Local & $O(N \textrm{log}N)$ & 0.618 & 0.619 & \textbf{0.902} \\
Community embedding (Louvain) \cite{faqeeh2018characterizing} & non-metric & Python & Local & $O(N \textrm{log}N)$ & 0.561 & 0.454 & \textbf{0.914} \\
\end{tabular}
\end{ruledtabular}
\end{table*}
\endgroup

\subsection{Mapping accuracy}

A general principle respected by all the embedding methods is that proximity in the embedding space is representative for similarity or proximity in the original graph. 
Indeed, some embedding methods work by directly finding the embedding configuration that best preserves pairwise distance or other similarity relationships.
For example, Isomap, Poincar\'e maps, and Hydra aim at preserving the shortest path distance among all pairs of nodes in the embedding space; Node2vec and HOPE try to encode certain similarity information.
Other methods follow the principle implicitly by fitting the observed network against proximity-preserving network models (see Sec.~\ref{sec:emb_methods} for details). 

A natural way to assess the quality of a method is to measure how accurately the embedding method maps nodes in the space so that pairwise graph proximity is preserved in the embedding.
We quantify the mapping accuracy of an embedding method in terms of the Spearman's correlation coefficient $\rho$ between the pairwise shortest path distance in the network and the pairwise distance in the embedding space. Note that it is infeasible to consider every possible pair of nodes for large networks. We therefore use a maximum of $10^5$ random pairs of nodes to approximate the Spearman's $\rho$ in case the total number of node pairs $N(N-1)/2>10^5$.

\begin{figure*}[!htb]
\includegraphics[width=0.95\textwidth]{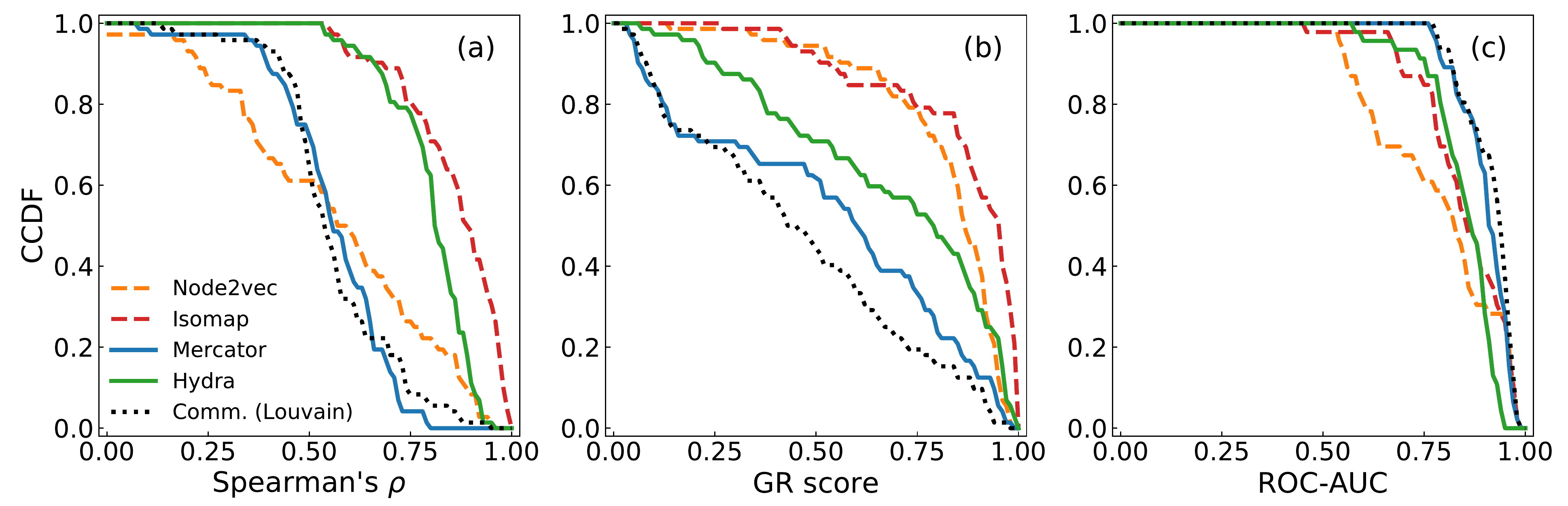}
\caption{\label{fig:comparison_ccdf_fix_parameter_all} \textbf{Aggregate performance in downstream tasks.} 
We show the complementary cumulative distribution function (CCDF) of (a) the Spearman's correlation coefficient of the mapping accuracy, (b) the GR score of greedy routing, and (c) the ROC-AUC scores of link prediction for different embedding methods on real-world networks. 
The average performance over all networks of an embedding on a task is equal to the area under the curve of the corresponding CCDF.
Since most of the embedding methods are stochastic, the data points in the figure are obtained by averaging the results from five independent repetitions.}
\end{figure*}

As mentioned above, we calculate the mapping accuracy of different embedding methods on 72 real-world networks. For sake of clarity, in Fig.~\ref{fig:comparison_ccdf_fix_parameter_all}(a), we plot the CCDF for some selected methods only.
The CCDF-AUC values of all embedding methods are listed in Table~\ref{tab:embedding_methods_summarize}.
Overall, we find that all methods do a good job in preserving graph proximity in the embedding space.

Isomap and Hydra top the ranking on this task. The finding is not surprising given that both methods aim at optimizing the congruence between pairwise proximity of nodes in the graph and in the embedding space. 
The mapping accuracy of Poincar\'e maps is not as high even though it also aims at preserving the shortest distance among pairs of nodes.
An advantage of Isomap and Hydra is that they can perform embedding in arbitrarily high-dimensional spaces, while Poincar\'e maps can only work in the two-dimensional hyperbolic space. Our experiments show that the mapping accuracy of Isomap and Hydra increases as the embedding dimension increases.
The results of Fig.~\ref{fig:comparison_ccdf_fix_parameter_all}(a) and Table~\ref{tab:embedding_methods_summarize} are obtained with $d=128$.
By setting $d=2$, Poincar\'e maps achieves the best performance; the performance of Hydra is also better than that of Isomap.
The main reason is that the two-dimensional Euclidean space may not be large enough to properly embed large networks (see \SM for details).

\subsection{Greedy routing}

Network embeddings may be used in greedy routing protocols devised for efficient network navigation~\cite{kleinberg2000navigation,boguna2009navigability}.
The task regards the delivery of a packet from a source node $s$ to a target node $t$.
The packet performs hops on the network edges, moving from one node to one of its neighbors at each stage of the navigation process.
In particular, according to the greedy protocol, at every stage of the process the packet moves to the neighbor that is closest to target $t$ according to a metric of distance.
Such a metric of distance is computed using 
knowledge about the embedding space and the nodes' coordinates.
If the packet reaches the target node $t$, the delivery is considered successful.
However, if the packet visits the same node twice, the delivery fails.
A good embedding for this task should be able to allow a high rate of successful deliveries along delivery paths that are not much longer than the true shortest paths.

In this work, we follow the literature and use the greedy routing score (GR score) 
to measure the performance of different embeddings in greedy routing~\cite{muscoloni2017machine}.
The GR score is defined as 

\begin{equation}
    \textrm{GR score} = \frac{2}{N(N-1)} \sum_{i > j} \frac{D_{ij}}{R_{ij}} \; ,
\label{eq:gr}
\end{equation}
where $D_{ij}$ is the shortest path length between nodes $i$ and $j$ in the original network, and $R_{ij}$ is the length of the actual delivery path followed by the packet according to the greedy routing protocol. All pairs of nodes are considered in the sum of Eq.~(\ref{eq:gr}), including 
those leading to successful and unsuccessful deliveries. 
For an unsuccessful delivery, $R_{ij}$ is infinite and $D_{ij}/R_{ij} = 0$. 
For a successful delivery along one of the shortest paths connecting $i$ to $j$, we have $D_{ij}/R_{ij} = 1$.
The GR score is 0 when all the deliveries are unsuccessful. The GR score equals 1 
when all packets are successfully delivered along the shortest path in the original network.
Note that it is impossible to test every pair of source-target nodes for large networks. In our experiments, we randomly select $10^4$ source-target pairs to approximate the GR score in case the total number of node pairs $N(N-1)/2>10^4$.

We show the CCDF of the GR scores for selected embedding methods in Fig.~\ref{fig:comparison_ccdf_fix_parameter_all}(b) and the CCDF-AUC values for all methods in Table~\ref{tab:embedding_methods_summarize}.
We note that all methods can facilitate network navigation to some extent. In general, there is 
a non-trivial
relationship between the performance in mapping accuracy and the one in greedy routing.
It is already known that Isomap performs well in this task~\cite{zhang2021model}. The relatively good performance of Node2vec is instead a new result. In part, the result can be explained by considering that embeddings obtained by Node2vec are based on the exploration of graph paths, a process that well informs a greedy navigation protocol. On the other hand, it seems that Euclidean-space-based embeddings better suit for this task than embedding methods relying on hyperbolic geometry and non-metric spaces. 
A possible explanation of our finding is that many of the non-Euclidean embedding methods focus on preserving local network properties rather than global ones. The only exception to this rule is Hydra, which in fact displays relatively higher performance than that of the other hyperbolic embedding methods. 

\subsection{Link prediction}

Link prediction is a standard task to evaluate the performance of graph embedding methods~\cite{liben2007link,goyal2018graph}.
The goal is predicting the existence or the non existence of edges between non-observed pairs of nodes.
There are potentially many different ways to implement the task.
In our case, we first remove $30\%$ randomly chosen edges from the original network while ensuring 
that the remaining graph is still formed by a single connected component.
The removed edges are used as the positive test set.
Then, we randomly sample a negative test set of non-existent edges with size identical to that of the positive test set.
The remaining network is fed to the embedding methods.
For each pair of nodes, the closer they are in the embedding space, the more likely they are connected.

The ability of an embedding to distinguish the edges from the positive and negative sets is measured by the area under the receiver-operating characteristic curve (ROC-AUC).
The ROC-AUC score ranges from 0.5 to 1. For perfect prediction, the ROC-AUC score equals to 1. The score is 0.5 for random guesses.
For small networks, removing $30\%$ of the edges may substantially distort the network structure and the link prediction results.
Therefore, we only consider real-world networks with more than 300 nodes for the link prediction task in this paper.
We show the CCDF of ROC-AUC scores for selected embedding methods in Fig.~\ref{fig:comparison_ccdf_fix_parameter_all}(c) and report the CCDF-AUC values for all methods in Table~\ref{tab:embedding_methods_summarize} as before.
All embedding methods yield comparable performance in this task.
Mercator and the community-based methods yield a slightly better performance than the other methods. The result can be a reflection of the fact that the embeddings are obtained by fitting graphs against probability laws for network connections, which immediately provide predictions for missing links. 
We also measure the area under the precision-recall curve (AUPR) for each method in the link prediction task, the results are qualitatively similar (see \SM for details).

\subsection{Embedding performance on synthetic networks}

In order to systematically analyze the performance of the different embedding methods, we also use 34 instances of synthetic networks generated by five types of network models: the popularity-similarity-optimization (PSO) model~\cite{papadopoulos2012popularity, papadopoulos2015network_1}, the Lancichinetti-Fortunato-Radicchi (LFR) model~\cite{lancichinetti2008benchmark}, the configuration model with power-law degree distribution and Poisson degree distribution (power-law networks and Poisson networks), and the model for spatial networks by Daqing {\it et al.}~\cite{daqing2011dimension} (see Sec.~\ref{sec:networks} for details of network models and parameters used). 

\begingroup
\squeezetable
\begin{table*}[!htb]
\caption{
\textbf{Embedding performance on synthetic networks.}
We summarize all the results obtained by the different embedding methods on  the synthetic network models  considered in this paper (i.e., PSO models, LFR networks, power-law networks, spatial networks, and Poisson networks).
From left to right, we report: name of the method, the CCDF-AUC of mapping accuracy on the various network models, the CCDF-AUC of greedy routing score on the same set of network models, the CCDF-AUC of link prediction ROC-AUC on the same set of network models.
Link prediction results for Poisson networks are excluded since no meaningful prediction can be made for the edges of random and homogeneous networks. 
See details about synthetic networks in Sec.~\ref{sec:networks}.
We highlight in bold face the top three methods for each network model and task combination.
Some values for Mercator and HyperLink are marked with * because the methods are not able to embed several networks. The results are qualitatively similar if we restrict the analysis on the subset of networks that all methods can process.
}
\label{tab:synthetic_results}
\begin{ruledtabular}
\begin{tabular}{r|ccccc|ccccc|cccc}
 & \multicolumn{5}{c|}{Mapping accuracy} & \multicolumn{5}{c|}{Greedy routing} & \multicolumn{4}{c}{Link prediction} \\
\hline
\textrm{Method} & \textrm{PSO} & \textrm{LFR} & \textrm{power-law} & \textrm{spatial} & \textrm{Poisson} & \textrm{PSO} & \textrm{LFR} & \textrm{power-law} & \textrm{spatial} & \textrm{Poisson} & \textrm{PSO} & \textrm{LFR} & \textrm{power-law} & \textrm{spatial}\\
\hline
Node2vec & 0.710 & \textbf{0.626} & \textbf{0.692} & \textbf{0.692} & \textbf{0.578} & \textbf{0.892} & \textbf{0.886} & \textbf{0.925} & \textbf{0.903} & \textbf{0.876} & 0.825 & 0.674 & 0.491 & \textbf{0.770} \\
HOPE & \textbf{0.740} & 0.444 & 0.662 & 0.547 & 0.442 & 0.742 & \textbf{0.740} & \textbf{0.873} & \textbf{0.775} & \textbf{0.768} & 0.750 & 0.678 & 0.523 & 0.697 \\
LE & 0.540 & 0.462 & 0.523 & 0.485 & 0.452 & 0.785 & 0.641 & 0.673 & 0.662 & 0.692 & 0.762 & 0.662 & 0.607 & 0.618 \\
Isomap & \textbf{0.943} & \textbf{0.789} & \textbf{0.853} & \textbf{0.863} & \textbf{0.652} & \textbf{0.872} & \textbf{0.846} & \textbf{0.887} & \textbf{0.885} & \textbf{0.794} & 0.818 & \textbf{0.729} & \textbf{0.647} & \textbf{0.733} \\
\hline
HyperMap & 0.379 & 0.314 & 0.365 & 0.283 & 0.266 & \textbf{0.797} & 0.265 & 0.528 & 0.371 & 0.294 & \textbf{0.848} & 0.695 & \textbf{0.653} & 0.660 \\
Mercator & 0.459* & 0.384 & 0.375 & 0.450 & 0.339 & 0.607* & 0.198 & 0.253 & 0.298 & 0.192 & 0.847* & 0.698 & 0.623 & 0.687 \\
Poincar\'e maps & 0.618 & 0.379 & 0.412 & 0.489 & 0.315 & 0.577 & 0.256 & 0.228 & 0.418 & 0.218 & 0.808 & 0.672 & 0.590 & 0.680 \\
HyperLink & 0.303 & 0.375 & 0.370 & 0.345 & 0.317* & 0.593 & 0.295 & 0.313 & 0.355 & 0.233* & 0.742 & 0.719 & 0.642 & 0.662 \\
Hydra & \textbf{0.898} & \textbf{0.666} & \textbf{0.773} & \textbf{0.685} & \textbf{0.528} & 0.765 & 0.371 & 0.574 & 0.422 & 0.480 & 0.783 & 0.671 & \textbf{0.663} & 0.632 \\
\hline
Comm. (Infomap) & 0.586 & 0.434 & 0.402 & 0.437 & 0.329 & 0.743 & 0.318 & 0.473 & 0.442 & 0.360 & \textbf{0.883} & \textbf{0.735} & 0.633 & \textbf{0.738} \\
Comm. (Louvain) & 0.543 & 0.384 & 0.353 & 0.388 & 0.309 & 0.592 & 0.178 & 0.185 & 0.203 & 0.149 & \textbf{0.883} & \textbf{0.740} & 0.638 & 0.732 \\
\hline

\end{tabular}
\end{ruledtabular}
\end{table*}
\endgroup

We apply the embedding methods to the synthetic networks, repeat the evaluation on three downstream tasks and report the performance in Table~\ref{tab:synthetic_results}.
We can see that the results on the synthetic network models are consistent with the results obtained on the real-world networks.
Isomap and Hydra are the top two methods for mapping accuracy.
Euclidean embeddings such as Node2vec and Isomap perform better than hyperbolic and community-based embeddings on greedy routing, while hyperbolic and community-based embeddings outperform Euclidean-based embedding methods on link prediction. 

\begin{table}[!htb]
\caption{
Synthetic network models considered in our analysis together with the corresponding network characteristics varied in our tests.
}
\label{tab:models_characteristics}
\begin{ruledtabular}
\begin{tabular}{r|l}
\textrm{Network model} & \textrm{Characteristic} \\
\hline
PSO model~\cite{papadopoulos2012popularity, papadopoulos2015network_1} & Clustering coefficient \\
LFR model~\cite{lancichinetti2008benchmark} & Modularity \\
Poisson network & Average degree \\
power-law network & Power-law exponent \\
spatial networks~\cite{daqing2011dimension} & Power-law exponent 
\end{tabular}
\end{ruledtabular}
\end{table}

By tuning the parameters of the network models, we can further study the effect of network characteristics on the performance of different embedding methods. The network models and the corresponding network characteristics analyzed in this paper are listed in Table~\ref{tab:models_characteristics}.

We find that certain network characteristics have strong effects on downstream tasks as follows:
(1) the ability of embedding methods to preserve graph distance deteriorates as the density of the network grows; (2) the ability of embedding methods to inform the greedy routing protocol improves as the network clustering coefficient increases but its modularity decreases; (3) the ability of embedding methods in inferring links between non-observed pairs of nodes improves as the network clustering coefficient increases, the network modularity grows, and the heterogeneity of the degree distribution increases.
Detailed results can be found in the \SM.
These effects are universal across different methods with a few exceptions.
For example, Isomap and Node2vec perform well in greedy routing regardless of the network characteristics.

\subsection{Summary of the results}

\begin{figure}[!htb]
\includegraphics[width=0.5\textwidth]{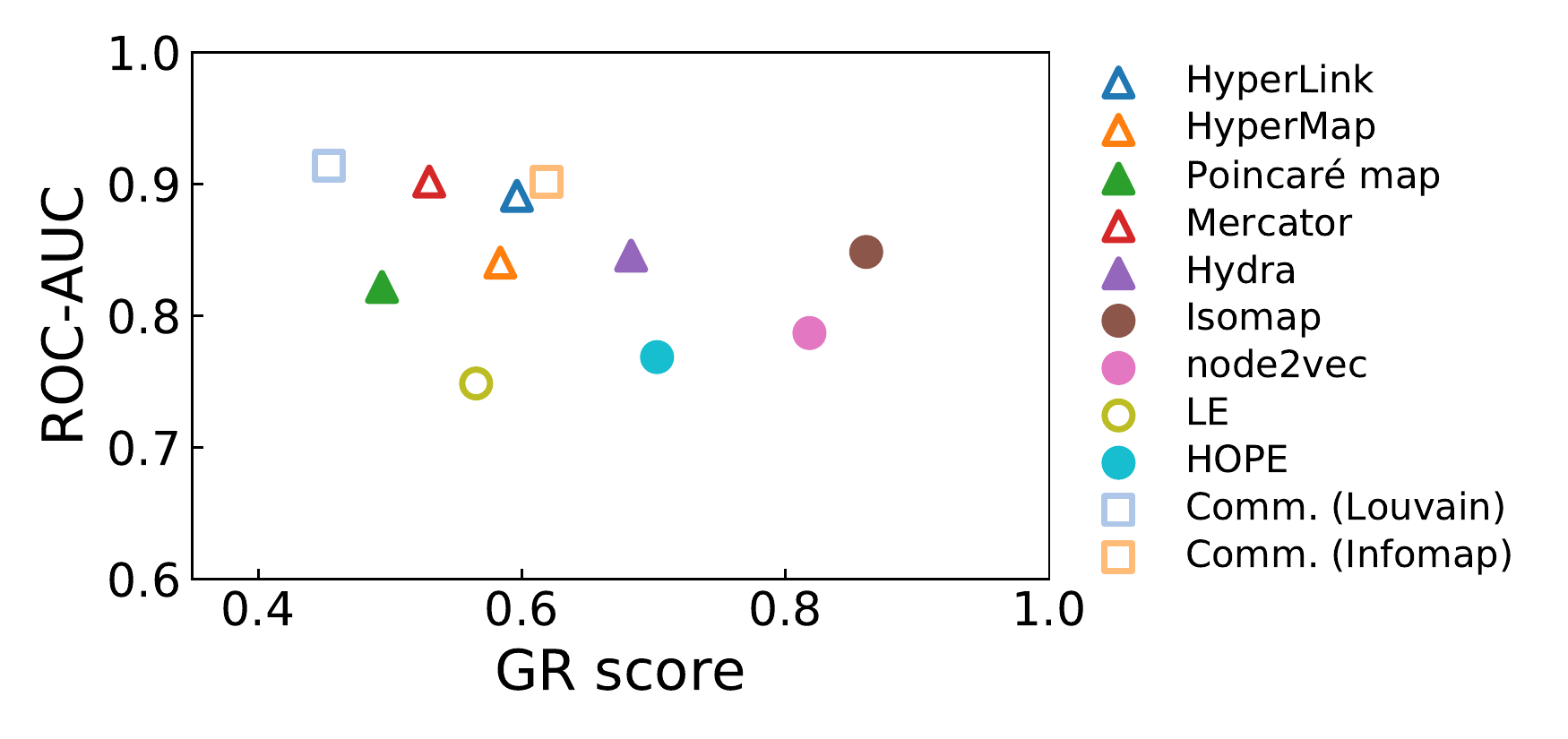}
\caption{\label{fig:methods_GR_LP} 
\textbf{Average performance in link prediction and greedy routing over a large corpus of real-world networks.}
We summarize here the same results as of Table~\ref{tab:embedding_methods_summarize}. We plot 
the CCDF-AUC values of ROC-AUC scores and GR scores for different embedding methods. 
Circles, triangles and squares represent Euclidean-, hyperbolic- and community-based embedding methods, respectively.
The hollow and solid symbols represent methods that preserve local and global network structural information, respectively.}
\end{figure}

To provide an overview of the performance of different embedding methods, we focus on link prediction and greedy routing, and summarize the results in Fig.~\ref{fig:methods_GR_LP}. 
The same analysis for synthetic networks can be found in the \SM.
We can see that Isomap and Node2vec outperform the other methods in greedy routing while community embedding, Mercator, and HyperLink yield better performance in link prediction.
However, no single method outperforms all the other methods in both tasks according to Fig.~\ref{fig:methods_GR_LP}.

We remark that the two tasks are fundamentally different, as link prediction is a local prediction task while greedy routing is a global discovery task.
Also, the position of an embedding method in the performance diagram shown in Fig.~\ref{fig:methods_GR_LP} seems partially predictable based on the type of space targeted by the embedding method and/or the type of network structural information that the method is able to preserve (see Table~\ref{tab:embedding_methods_summarize}). 
As a general rule of thumb, methods that preserve local information excel in link prediction, and algorithms that preserve global structure achieve optimal performance in greedy routing.

\begin{figure}[!htb]
\includegraphics[width=0.5\textwidth]{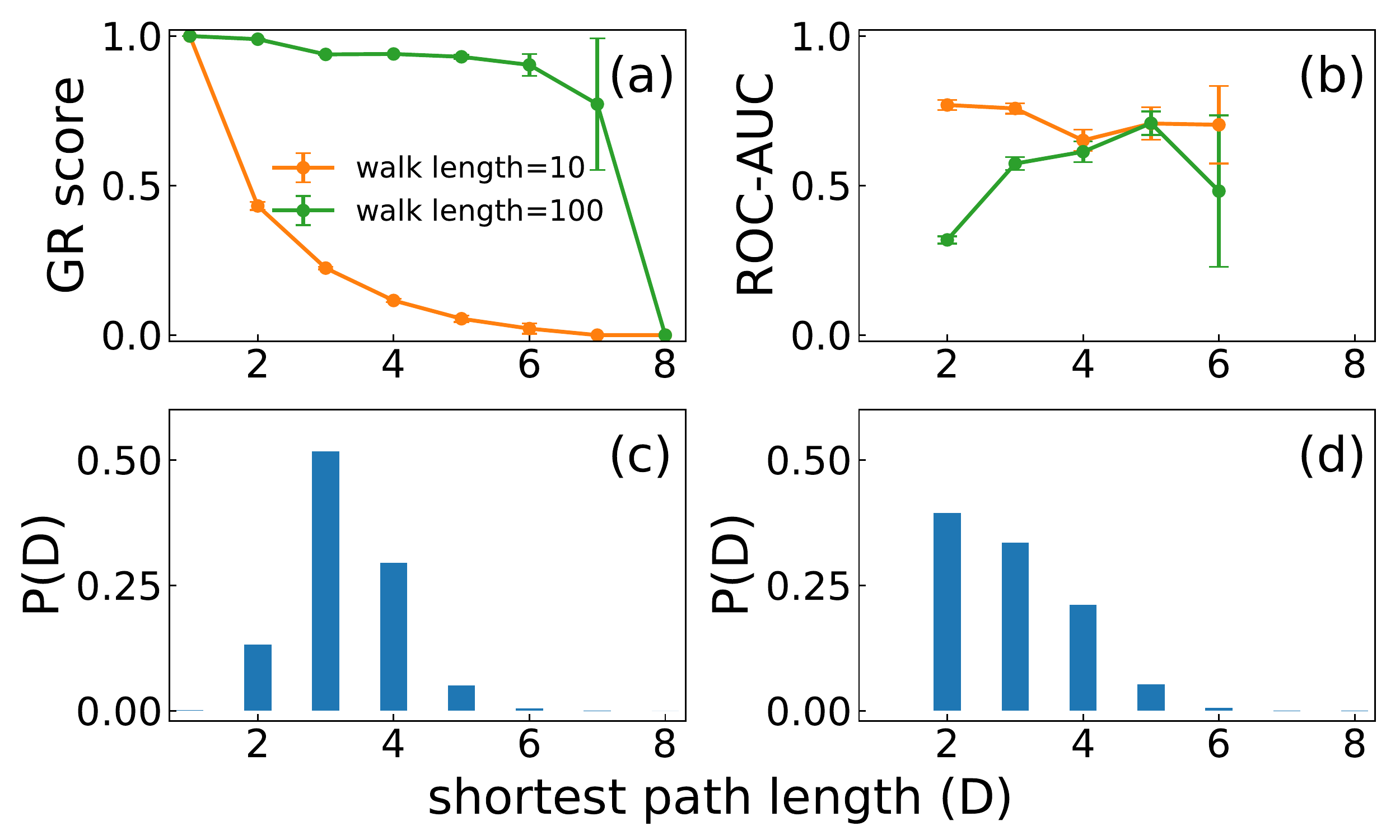}
\caption{\label{fig:gr_lp_detail_IPv6} 
\textbf{Greedy routing and link prediction results obtained by Node2vec with different walk length on the IPv6 Internet network.}
We display (a) the relation between GR score and the shortest path length between node pairs involved when using Node2vec with different walk length ($l = 10$ and $l = 100$) to guide greedy routing, (b) same as (a), but for ROC-AUC scores in link prediction, (c) the distribution of distance between node pairs involved in greedy routing, and (d) same as (c), but for link prediction.
The data points in the figure are obtained by averaging the results of 10 experiments, the error bars indicate the standard deviation. 
}
\end{figure}

To further test our hypothesis, we take advantage of Node2vec.
The algorithm acquires structural information by means of random walks with restart. The length of the random walks serves as a proxy for the typical scale of structural information that is preserved by the embedding. 
We apply Node2vec with walk length $l=10$ and $l=100$ on the Ipv6 Internet network~\cite{Kleineberg2016Hidden} and use the resulting embeddings to perform greedy routing and link prediction.
Instead of reporting the overall performance, we group the node pairs involved in the tasks by their shortest path distance in the network and then calculate the scores within each group. For $l=10$, the GR score decreases quickly as the distance between source and target nodes increases. The performance for $l=100$ in greedy routing is instead almost unaffected by the 
source-to-target distance. Performance in link prediction obtained for $l=10$ is far better than the one obtained for $l=100$. We note that the vast majority of links tested have distance $D=2$, which corresponds to the maximum gap in performance between the embeddings obtained for $l=10$ and $l=100$.

\section{Computational complexity and running time}

\begin{figure}[!htb]
\includegraphics[width=0.5\textwidth]{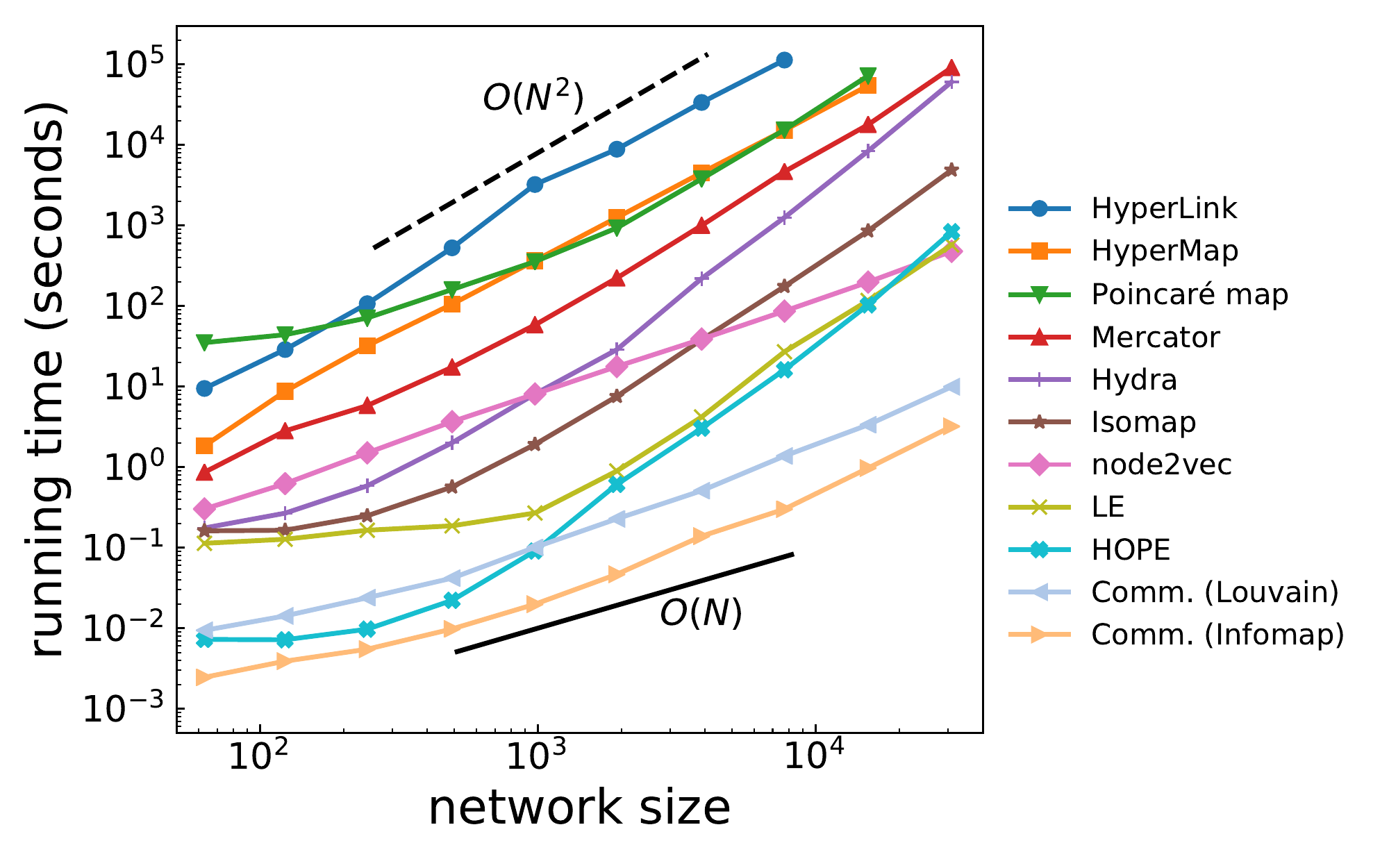}
\caption{\label{fig:running_time}
\textbf{Running time vs. network size.} 
We show the running time of different embedding methods in relation to the size of PSO models.
The network size ranges from $N=2^6$ to $N=2^{15}$. Other parameters of the PSO models are: average degree $\langle k \rangle = 5$, power-law exponent $\gamma = 2.1$, temperature $T = 0.5$. Each data point is the average of five simulations. For HyperMap, we use the hybrid algorithm without correction steps and enable the speedup mode by setting $k_{\textrm{speedup}} = 10$ (see Sec.~\ref{sec:emb_methods} for details). The black full line indicates linear scaling; the black dashed line denotes quadratic scaling.}
\end{figure}

\begingroup
\squeezetable
\begin{table}[!htb]
\caption{
\textbf{Node2vec and community embedding on large networks.}
We report the performance on mapping accuracy (Spearman's $\rho$), greedy routing (GR score), and link prediction (ROC-AUC score) as well as the running time (seconds) of Node2vec and community embeddings with Infomap and Louvain algorithms on the YouTube friend network (N = 1,134,890) and the AS Skitter network (N = 1,694,616).
}
\label{tab:large_networks}
\begin{ruledtabular}
\begin{tabular}{l|l|ccc}
\textrm{Network} & \textrm{Metric} & \textrm{Node2vec} & \textrm{Infomap} & \textrm{Louvain} \\
\hline
\multirow{4}{*}{YouTube friend} & Mapping accuracy & 0.620 & 0.499 & 0.352 \\
                                & Greedy routing & 0.478 & 0.071 & 0.588 \\
                                & Link prediction & 0.959 & 0.962 & 0.976 \\
                                & Running time & 33,045 s  & 4,938 s & 732 s\\
\hline
\multirow{4}{*}{AS Skitter} & Mapping accuracy & 0.582 & 0.403 & 0.033 \\
                            & Greedy routing & 0.348 & 0.117 & 0.363 \\
                            & Link prediction & 0.998 & 0.991 & 0.983 \\
                            & Running time & 85,356 s & 3,149 s & 1,895 s \\
\end{tabular}
\end{ruledtabular}
\end{table}
\endgroup

Scalability is another important factor when choosing the proper embedding method.
We summarize the computational complexity in Table~\ref{tab:embedding_methods_summarize}.
Hyperbolic embedding methods have $O(N^2)$ computational complexity at least, while Euclidean and non-metric methods often scale linearly with the system size.

To directly compare the running time of the various embedding techniques, we
apply all the methods to a series of networks with different sizes generated by the popularity-similarity-optimization (PSO) model~\cite{papadopoulos2012popularity, papadopoulos2015network_1}.
All the experiments are performed on a server equipped with Intel Xeon Platinum 8268 CPUs (2.90GHz) and 1.5TB RAM.
Although the server have multiple processors, all the methods are allowed to use one processor only.
Figure~\ref{fig:running_time} shows the relation between the running time and the network size for all the embedding methods.
The results confirm that the Euclidean and the non-metric embedding methods tend to be much faster than the hyperbolic embedding methods.
When we apply the embedding algorithms to different network models and measure their computational time, results are qualitatively similar.

Among the methods tested, only Node2vec and community embedding methods (both variants with Louvain~\cite{blondel2008fast} and Infomap~\cite{rosvall2008maps}) can easily scale up to large networks.
As a demonstration, we apply them to two real-world networks with more than one million nodes.
They complete the embedding in 
about
24 hours and 1.4 hours, respectively, without compromising the performance on downstream tasks (see details in Table~\ref{tab:large_networks}).
In order to avoid unnecessary memory and time usage while applying Node2vec on networks with millions of nodes, we use a program optimized for unweighted networks and specific algorithm parameter values ($p = 1$ and $q = 1$). 

In our experiments, we try to use the implementation shared by the creators whenever possible.
For classic methods such as LE and Isomap, we use the implementation provided by the Python package scikit-learn~\cite{scikit-learn}.
We implement Node2vec and community embedding in Python with the help of some open source packages.
Note that this is not the ideal setup for comparing the running time of different methods since the programming language (see Table~\ref{tab:embedding_methods_summarize}) used 
can heavily affect the results and the implementation used in our experiments can sometimes be further optimized.
Instead, our experiments mimic a more practical scenario where practitioners hope to quickly apply the embedding methods without spending too much time improving or even implementing the methods themselves.
The results here provide a rough estimation of the expected running time when using the most accessible implementation.

\section{Discussion}

In this work, we consider a large corpus of real-world and synthetic networks, and measure the performance of several 
embedding methods in solving specific network tasks. 
We find that Isomap and Node2vec outperform the other methods in greedy routing.
As for link prediction, community embedding, Mercator, and HyperLink all yield excellent performance.
Our results on synthetic network models indicate that type and feature of the target networks are not important when choosing the embedding method.
Instead, one possible principle is that the methods aim at preserving global network structure excel in greedy routing, and methods only capturing local information achieve optimal performance in link prediction.
Also, our analyses of the algorithm running time show that hyperbolic methods are much slower than other methods, suggesting that they are not yet well suited for embedding large-scale networks. 

We stress that not all factors that are important in the decision of using an embedding method instead of another are measurable and quantifiable.
Some methods may provide valuable insights into the characteristics of networks although their performance may not be comparable with that of others in certain tasks.
For practical tasks, many other features may also be crucial.
A method can be chosen because its implementation is easy to access and configure, and the method can process different input networks.
For instance, we had to exclude some embedding methods from our experiments because we were unable to find adequate implementations. 
Also, some of the methods considered in this paper require proper calibration of input parameters to be successful in downstream tasks~\cite{tandon2021community}. Calibration is generally a computationally expensive operation, and there may be practical situations where calibration can not even be performed. 

All things considered, we believe that the Euclidean embedding methods like Node2vec and Isomap should be the first options for practitioners since they have stable and widely available implementations, and they yield decent performance in all tasks. The non-Euclidean embedding methods still present some challenges.
Their non-Euclidean nature makes it non-trivial to incorporate their results to common downstream tasks in general, which may limit their applicability. 
Nevertheless, the fact that the non-Euclidean methods stand out in certain tasks suggests that they have a great potential, calling for further investigation and improvement.

\section{Acknowledgements}

Y.-J.Z. acknowledges support from China Scholarship Council (No.201906180029).
Y.-J.Z. and F.R. acknowledge partial support from the National Science Foundation (Grant No. CMMI-1552487).
F.R. acknowledges partial support from the US Army Research Office (Grant No. W911NF-21-1-0194).

\section{Methods}

\subsection{Network Embedding Methods}
\label{sec:emb_methods}

Network embedding methods are sets of procedures that map the nodes of the input network into points in the target space.
The coordinates of the nodes serve as the vector representation of the networks and the pairwise distance of different nodes correspond to their proximity or similarity in the input networks.
Depending on the target spaces, the representation of the embedded network and the definition of the distance between nodes in the embedding space vary.
Here we group different embedding methods by their target spaces, i.e., Euclidean, hyperbolic, and non-metric spaces.

\subsubsection{Euclidean embedding methods}

For Euclidean embedding methods, each node $i$ can be described by a $d$-dimensional vector $\textbf{x}_i = (x_i^{(1)}, ..., x_i^{(d)})$ where $d$ is the space dimension and 
serves as a free parameter for all Euclidean embedding methods.
There are several ways to calculate the distance between two nodes in Euclidean embedding space. The most common two, Euclidean distance and dot product, are used in this work. The Euclidean distance between node $i$ and $j$ is defined as
\begin{equation}
    \textrm{dist}_{ij} = \lVert \textbf{x}_i - \textbf{x}_j \lVert  = \sqrt{\sum_{v=1}^d (x_i^{(v)} - x_j^{(v)})^2} \; . 
    \label{eq:Euclidean_distance}
\end{equation}
The dot product between node $i$ and $j$ is given by
\begin{equation}
    \textbf{x}_i \cdot \textbf{x}_j = \sum_{v=1}^d x_i^{(v)} x_j^{(v)} \; .
    \label{eq:dot_product}
\end{equation}
Note that the similarity between two vectors is proportional to their dot product.
So we use
\begin{equation}
    \textrm{dist}_{ij} = -\textbf{x}_i \cdot \textbf{x}_j \; ,
    \label{eq:dist_dot_product}
\end{equation}
as effective distance in the dot product approach.

Node2vec, LE, HOPE and Isomap are the four Euclidean embedding methods we consider in this paper.
We use either the distance of Eq.~(\ref{eq:Euclidean_distance}) or Eq.~(\ref{eq:dist_dot_product}) depending on the objective function that a method minimizes and the actual downstream task.
Eq.~(\ref{eq:Euclidean_distance}) is used for LE and Isomap in this paper.
For Node2vec and HOPE, we use Eq.~(\ref{eq:dist_dot_product}) for link prediction according to their objective functions, and Eq.~(\ref{eq:Euclidean_distance}) for mapping accuracy and greedy routing because it yields much better performance than when distance is calculated according to Eq.~(\ref{eq:dist_dot_product}) (see \SM for details).

Next, we briefly introduce each method and the parameters used in our experiments.

\begin{itemize}
    \item[(1)] 
    \textit{Node2vec}~\cite{grover2016node2vec}: 
    Node2vec first generates multiple node sequences using random walks with fixed length, then finds the vector representations that maximize the probability of co-occurrence of the nodes in the sequences.
    There are some tunable parameters for Node2vec, such as walk length $l$, window size, the bias parameters of the random walk dynamics 
    $p$ and $q$, and the embedding dimension $d$. In this work, we use the default setting: window size $ = 10$, $p = 1$ and $q = 1$.
    
    We find that the walk length can greatly affect different downstream tasks. 
    The main reason is that walk length directly control the type of information that the resulting embedding preserves. Short walk lengths preserve local structural information; long walk lengths preserve global structure. As expected,
    according to our tests on several real-world networks, increasing the walk length improves the performance of mapping accuracy and greedy routing, but worsens link prediction (see \SM for details). So we 
    set $l=10$ for link prediction and $l=100$ for the other two tasks in this paper.
    
    In general, the larger the dimension $d$, the better the embedding. But for Node2vec, the performance in downstream tasks may decrease slightly as $d$ increases (see \SM). In this work, we set 
    $d = \min \{N, 128\}$
    for all embedding methods that can work with high ($d>2$) dimensional embedding space, which is considered
    a sufficiently high value to achieve nearly-optimal embeddings of networks~\cite{gu2020defining}.
    We make this choice to maintain the simplicity of the experiments without introducing strong biases towards certain methods.

    \item[(2)] \textit{Laplacian eigenmaps (LE)}~\cite{belkin2001laplacian}: 
    LE aims to place the nodes that are connected with each other closely in the embedding space by minimizing the objective function
    \begin{equation}
        E_{\textrm{LE}} = \sum_{ij} \lVert \textbf{x}_i - \textbf{x}_j \lVert^2 A_{ij} = tr(\textbf{X}^T \textbf{L} \textbf{X}) \; ,
    \end{equation} 
    where $\textbf{X} = (\textbf{x}_1, \textbf{x}_2, ... \textbf{x}_n)^T$ is the low-dimensional representation matrix of the network, $\mathbf{A}$ is the  adjacency matrix of the network ($A_{ij} =A_{ji} = 1$ if nodes $i$ and $j$ are connected, otherwise $A_{ij} = A_{ji} =0$), $\textbf{L} = \textbf{K} - \textbf{A}$ is the Laplacian matrix and $\textbf{K}$ is the diagonal matrix with $K_{ii} = \sum_{j} A_{ji}$.
    LE further requires $\textbf{X}^T \textbf{K} \textbf{X} = \textbf{I}$ to eliminate trivial solutions. 
    To obtain a $d$-dimensional embedding, one can simply extract the eigenvectors that correspond to the $d$ smallest non-zero eigenvalues of the solution to $\textbf{Lx} = \lambda \textbf{Kx}$.
    
    LE only has one tunable parameter: dimension $d$. We set it to $d = \min \{N, 128\}$.
    
    \item[(3)] \textit{HOPE}~\cite{ou2016asymmetric}: 
    Given a node similarity definition, HOPE seeks to preserve the similarity matrix $\textbf{S}$ in the embedding space by minimizing 
    \begin{equation}
        E_{\textrm{HOPE}} =  \lVert \textbf{S} - \textbf{x} \textbf{x}^T \lVert ,
    \end{equation}
    through singular value decomposition (SVD).
    HOPE can work with different node similarity definitions; here we use Katz index, which is calculated by
    \begin{equation}
        \textbf{S}^{\text{Katz}} = 
        \beta \sum_{l=1}^{\infty} \, \textbf{A}^l \; ,
    \end{equation}
    where $\textbf{A}$ is the adjacency matrix of the network and $\beta$ is the decay parameter.
    HOPE requires $\beta < 1/\lambda_{max}$, with $\lambda_{max}$ principal eigenvalue of the matrix $\textbf{A}$ . We set $\beta = 1/\lambda_{max} - 0.001$ for all experiments. The embedding dimension $d$ is set to $d = \min \{N, 128\}$.
    
    \item[(4)] \textit{Isomap}~\cite{tenenbaum2000global}: 
    Isomap tries to preserve the shortest path distance between each pair of nodes. It first calculates the shortest path distance matrix \textbf{D} of a network. Then multidimensional scaling (MDS)~\cite{borg2005modern} is applied to \textbf{D} to obtain a \textit{d}-dimensional representation of the network that minimize the stress function
    \begin{equation}
        E_{\textrm{ISO}} =\sum_{ij} \left[ D_{ij} - \lVert \textbf{x}_i - \textbf{x}_j \lVert \right]^2 \; .
    \end{equation}
    We set the embedding dimension $d = \min \{N, 128\}$ for Isomap in all experiments.
    
\end{itemize}

\subsubsection{Hyperbolic embedding methods}
\label{sec:hyper_methods}

For hyperbolic embedding, nodes are usually considered as points on the Poincar\'e disk. Two coordinate systems are often used in the literature, i.e., the polar coordinates $(r, \theta)$ and the Poincar\'e coordinates $\textbf{y} = (y^{(1)}, y^{(2)})$.
The Poincar\'e coordinates are similar to the Euclidean coordinates but represent points in hyperbolic space.
They can also be extended to arbitrary dimension, i.e., $\textbf{y} = (y^{(1)}, ..., y^{(d)})$, to represent points in the Poincar\'e ball.

When using the polar coordinates, the distance between node $i$ and $j$ can be calculated by
\begin{equation}
    \textrm{dist}_{ij} = \textrm{arcosh}(\textrm{cosh}r_i \textrm{cosh}r_j - \textrm{sinh}r_i \textrm{sinh}r_j \textrm{cos}(\Delta \theta)) \; ,
    \label{eq:hyperbolic_distance}
\end{equation}
where $\Delta \theta = \pi - | \pi - |\theta_i - \theta_j||$ is the angle between the two nodes.

When using the Poincar\'e coordinates, the distance between node $i$ and $j$ can be calculated by
\begin{equation}
    \textrm{dist}_{ij} = \textrm{arcosh} \bigg(1 + 2\frac{\lVert \textbf{y}_i - \textbf{y}_j \rVert ^2}{(1 - \lVert \textbf{y}_i \rVert ^2)(1 - \lVert \textbf{y}_j \rVert ^2)}\bigg) \; .
    \label{eq:poincare_distance}
\end{equation}

The two-dimensional Poincar\'e coordinates $(y^{(1)}, y^{(2)})$ and the polar coordinates $(r, \theta)$ of hyperbolic space can be converted to each other by
\begin{equation}
    \begin{array}{ll}
        r &= 2 \textrm{artanh} (\sqrt{(y^{(1)})^2 + (y^{(2)})^2}) \; ,  \\
        \theta &= \textrm{atan2} (y^{(2)}, y^{(1)}) \; .
    \end{array}
\end{equation}

Among the hyperbolic embedding methods considered in this work, HyperMap, Mercator, and HyperLink use polar coordinates; Poincar\'e maps and Hydra use Poincar\'e coordinates.
Poincar\'e maps focus on the two-dimensional disk while Hydra can embed networks in higher-dimensional hyperbolic spaces.

We briefly introduce each method and the parameters used in our experiments in the following. 

\begin{itemize}
    \item[(1)] \textit{HyperMap}~\cite{papadopoulos2015network_1, papadopoulos2015network_2}: Popularity-similarity-optimization (PSO) model~\cite{papadopoulos2012popularity, papadopoulos2015network_1} is a growing network model that can simultaneously capture the heterogeneity degree distribution and strong clustering of real-world networks.
    Nodes of the PSO model are embedded in 
    the hyperbolic space and their coordinates have clear interpretations: the radial coordinate represent the node popularity, and the difference between angular coordinates of a node pair represents the similarity between them.
    The PSO model consists of a probability law for the existence of edges between pairs of nodes in the network depending on their distance in the hyperbolic space, i.e., Eq.~(\ref{eq:hyperbolic_distance}). 
    
    As an embedding method, HyperMap embeds an input network to the hyperbolic space by fitting the network against the PSO model. The fit is performed by maximizing the likelihood of observed edges according to the PSO connection probability law. As the maximum likelihood problem cannot be solved exactly, different variants of the HyperMap algorithm exploit different strategies to find approximate solutions. These variants include the link-based method~\cite{papadopoulos2015network_1}, the common-neighbors based method (also called HyperMap-CN)~\cite{papadopoulos2015network_2}, and the hybrid method~\cite{papadopoulos2015network_2} that uses the common-neighbors based method for high degree nodes and the link-based method for the rest of the nodes. 
    The computational complexity of the above-mentioned algorithms is at least $O(N^3)$.
    There is also a speed-up version of the hybrid method, which can reduce the computational complexity of the method down to $O(N^2)$ without compromising the embedding quality too much.
    
    In this paper, we use the speed-up version of HyperMap.
    This method has extra correction steps that can marginally improve the results but have a very high computational complexity so we disable them.
    It has a parameter $k_{\textrm{speedup}}$ to control the level of acceleration.
    We set $k_{\textrm{speedup}} = 10$ for networks with size $N < 10,000$ and $k_{\textrm{speedup}} = 40$ for networks with size $N > 10,000$. 
    
    The input parameters of HyperMap include the temperature $T \in [0, 1)$, which reflects the average clustering level of a network. A higher temperature means that the network is less clustered.
    Identifying the ideal temperature value for each network requires scanning the parameter space, which is infeasible in our experiments.
    Instead, we test the overall performance of HyperMap for different values of the temperature parameter
    on several real-world networks, and find that temperatures that are not too large nor too small generally yield decent performance.
    So, we set temperature $T = 0.5$ in all experiments.
    Another input parameter of HyperMap is the exponent $\gamma \ge 2$ of the power-law degree distribution of the  network.
    Note that not all real-world networks display a power-law degree distribution.
    To apply HyperMap to all the networks considered, we use the code shared by Broido {\it et al.}~\cite{broido2019scale} to estimate a suitable $\gamma$ value for every network. 
    If the estimated $\gamma$ value is smaller than $2.1$, we set $\gamma = 2.1$.
    
    \item[(2)] \textit{Mercator}~\cite{garcia2019mercator}: 
    Mercator learns the hyperbolic representations of networks by matching them with the $\mathbb{S}^1/\mathbb{H}^2$ model~\cite{serrano2008self, krioukov2010hyperbolic}. The $\mathbb{S}^1/\mathbb{H}^2$ model is the static version of the PSO model. While PSO model can only generate networks with pure power-law degree distribution, the $\mathbb{S}^1/\mathbb{H}^2$ model can generate networks with arbitrary degree distributions.
    Besides the input network itself, Mercator does not require any input parameters.
    
    \item[(3)] \textit{Poincar\'e maps}~\cite{klimovskaia2020poincare}:
    Poincar\'e maps aims to preserve the pairwise shortest path length just like Isomap. There are several free parameters of Poincar\'e maps.
    For example, the Gaussian kernel width $\sigma_P$ is related to the calculation of the global proximity of the original network, the scaling parameter $\gamma_P$ is used to tune the scattering of the embedding.
    We find that these parameters have little effect on the results. In this paper, we use the default setting $\sigma_P = 1$ and $\gamma_p = 2$ in all experiments. The maximum number of epochs for the embedding optimization is set to $e =1000$.
    
    \item[(4)] \textit{Hydra}~\cite{keller2020hydra}:
    Like Poincar\'e maps and Isomap, Hydra (HYperbolic Distance Recovery and Approximation) also seeks to preserve pairwise shortest path length.
    The difference between Poincar\'e maps and Hydra is that Hydra can work in hyperbolic spaces of arbitrary dimension, while Poincar\'e maps is designed for the two-dimensional space only. The dimension $d$ is the only one free parameter of Hydra. We set $d = \min \{N, 128\}$ in all experiments.
    
    \item[(5)] \textit{HyperLink}~\cite{kitsak2020link}:
    HyperLink is a model-based hyperbolic embedding method designed for link prediction. It tries to fit the networks to the random hyperbolic graphs (RHGs) model, which is equivalent to the $\mathbb{S}^1/\mathbb{H}^2$ model used in Mercator.
    HyperLink assumes that a fraction $p$ of links are missing when the embedding of the network is performed.
    In addition to $p$, other input parameters of HyperLink include the exponent $2 < \gamma < 3$ of the degree distribution, the temperature $T$, the number of layers $m$, and the coefficient $g$ that controls the size of the mesh in the angular space. In our experiments, we use the default settings $m = 20$ and $g = 1$. 
    We aid the method by setting $p=0.3$ in link prediction, and  $p=0$ in other tasks.
    The estimation of $\gamma$ is the same as in HyperMap. We set $\gamma = 2.1$ if the estimated $\gamma < 2.1$ and $\gamma = 2.9$ if the estimated $\gamma > 2.9$ in order to satisfy the requirement.
    Like HyperMap, the temperature $T$ is a free parameter for HyperLink.
    We test the overall performance of HyperLink for different $T$ values on some real-world networks, and find that $T = 0.3$ yields the best performance overall.
    Therefore, we set $T = 0.3$ in our experiments.

\end{itemize}

\subsubsection{Non-metric embedding method}

Community embedding~\cite{faqeeh2018characterizing} is a non-metric embedding method inspired by the analogy between hyperbolic embeddings and network community structure. 
It embeds networks using information about their community structures: node $i$ is represented by the coordinates $(k_i, \sigma_i)$, where $k_i$ is node's degree and $\sigma_i$ is the index of the community which the node belongs to.
There are many community detection algorithms available on the market. Here, we use two popular ones:  Infomap~\cite{rosvall2008maps} and Louvain~\cite{blondel2008fast}. 
After the community partition of a network is obtained, nodes in the same communities are merged together to generate supernodes, which then form a supernetwork.
The edge weight between community $a$ and $b$ in the supernetwork is defined as
\begin{equation}
    w_{ab} = 1 - \textrm{ln} \rho_{ab} \; , \textrm{if} \; \rho_{ab} > 0 \; ,
\end{equation}
and $w_{ab} =0$, otherwise. $\rho_{ab}$ is the ratio between the total number of edges between communities $a$ and $b$ and the sum of the node degrees in community $a$. 

The fitness between nodes $j$ and $i$
is defined as
\begin{equation}
    f_{ij} = \beta D_{\sigma_i \sigma_j} - (1 - \beta)\textrm{ln} \, k_i \; ,
    \label{eq:fitness}
\end{equation}
where $D_{\sigma_i \sigma_j}$ is the shortest path length between communities $\sigma_i$ and $\sigma_j$ in the supernetwork,  $k_i$ is the degree of node $i$, and $0 \le \beta \le 1$  is a free parameter. In order to maximize the overall performance of community embedding on different tasks, we test the effect of $\beta$ for the tasks on some real-world networks, and set $\beta = 0.3$ for all experiments.
Note that the fitness of Eq.~(\ref{eq:fitness}) is an asymmetric function, i.e., $f_{ij} \neq f_{ji}$.
In this paper we symmetrize it as
\begin{equation}
    \bar{f}_{ij} = \frac{f_{ij} + f_{ji}}{2},
\end{equation}
and we treat it at the same footing as of a distance between nodes $i$ and $j$, i.e.,
\begin{equation}
    \textrm{dist}_{ij} = \bar{f}_{ij} \: ,
\end{equation}
even though $\bar{f}_{ij}$ is not a proper metric of distance.

\subsection{Networks}
\label{sec:networks}

\begin{figure}[!htb]
\includegraphics[width=0.45\textwidth]{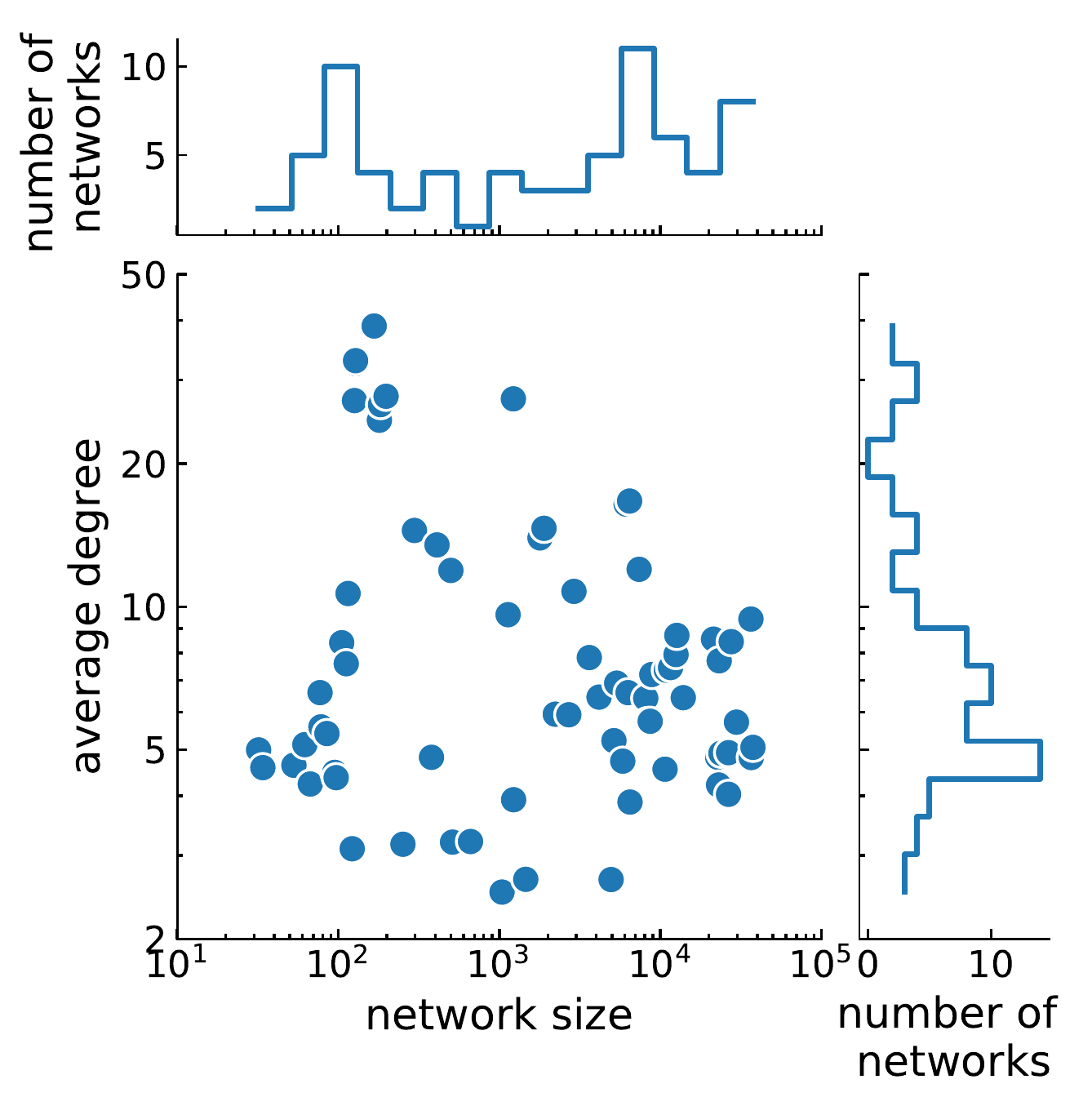}
\caption{\label{fig:real_world_networks} 
\textbf{Summary statistics of the real-world networks considered in this study.}
In the main panel, we show the scatter plot of the
average degree $\langle k \rangle$ versus network size $N$. 
Each point is a real network in our dataset.
Side panels are used to display 
non-normalized distributions of $\langle k \rangle$ and $N$.
}
\end{figure}

In this paper, we use both real-world networks and synthetic networks. 
All networks are unweighted and undirected.
We consider 72 real-world networks from different domains, including social, biological, technological, transportation, and Internet networks.
Sizes of these networks ranges from 32 to 37,542 nodes.
Figure~\ref{fig:real_world_networks} shows the average degree versus network size for all the 72 real-world networks used.
Two networks with more than one million nodes are also considered for Node2vec and community embedding particularly to demonstrate their scalability.
The full list of the real-world networks and some of their basic information can be found in the \SM.
Only the largest connected component of the various network is considered in our analysis.

We use 34 synthetic networks 
generated according to different models. 
We ensure that each network instance consists of one connected component only.
The network models considered are reported below.

\begin{itemize}
    \item[(1)] \textit{Popularity-similarity-optimization (PSO) model}~\cite{papadopoulos2012popularity, papadopoulos2015network_1}: PSO model grows networks by adding nodes to a hidden hyperbolic space.
    Nodes close with each others in the hidden space are then connected to form the edges.
    There are several parameters that could affect the properties of the generated networks: network size $N$, temperature $T$, average degree $\langle k \rangle$, and exponent $\gamma$ of the power-law degree distribution $P(k) \sim k^{-\gamma}$. Temperature $T\in [0, 1)$ controls the average clustering in the network, which is maximized at $T = 0$. We generate six instances of the PSO model with the following parameters: $N = \{1000; 10,000\}$, $T = \{0.1; 0.5; 0.9\}$, $\gamma = 2.1$, $\langle k \rangle = 5$.
    
    \item[(2)] \textit{Lancichinetti-Fortunato-Radicchi (LFR) model}~\cite{lancichinetti2008benchmark}: The LFR model generates networks with community structure, and both the degree distribution $P(k)$ and community size distribution $P(S)$ follow power-law distribution, i.e., $P(k) \sim k^{-\gamma}$ and $P(S) \sim S^{-\tau}$. Input parameters that are required to generate instances of the model are the network size $N$, the exponents $\gamma$ and $\tau$, the average degree $\langle k \rangle$, the maximum degree $k_{\textrm{max}}$, the minimum and maximum community size $c_{\textrm{min}}$ and $c_{\textrm{max}}$, and the mixing parameter $\mu$ that determines how strong the community structure is. A small value of $\mu$ corresponds to a strong community structure. We generate eight instances of LFR models, the parameters are $N = \{1000; 10,000\}$, $\mu = \{0.1; 0.3; 0.5; 0.7\}$, $\gamma = 2.1$, $\tau = 2$, $\langle k \rangle = 5$, $k_{\textrm{max}} = 50$, $c_{\textrm{min}} = 10$, $c_{\textrm{max}} = 0.1N$.
    
    \item[(3)] \textit{Poisson networks}: They are generated by feeding Poisson degree distributions to the configuration model. Two tunable parameters are the size of network $N$ and average degree $\langle k \rangle$. We use eight instances of Poisson networks with the following parameters: $N = \{1000; 10,000\}$, $\langle k \rangle = \{4; 6; 8; 10\}$.
    
    \item[(4)] \textit{Power-law networks}: They are generated by feeding power-law degree distributions to the configuration model. The tunable parameters are the network size $N$ and the power-law exponent $\gamma$. The average degree of a network can be controlled by setting the minimum value of the node degrees, namely $k_{\textrm{min}}$. We use six instances of power-law networks and the parameters are $N = \{1000; 10,000\}$, $\gamma = \{2.1; 2.5; 2.9\}$, $k_{\textrm{max}} = 100$. We use either $k_{\textrm{min}} = 2$ or $k_{\textrm{min}} =3$ for the various nodes in the network to ensure an average degree $\langle k \rangle \simeq 5$.
    
    \item[(5)] \textit{Spatial networks}~\cite{daqing2011dimension}: The model generate spatial networks that are embedded in two-dimensional regular lattice. Both the degree distribution $P(k)$ and the Euclidean distance distribution of edges $P(r)$ follow power-law distributions, i.e., $P(k) \sim k^{-\gamma}$ and $P(r) \sim r^{-\alpha}$. The input parameters of the model are the network size $N$, the exponents $\gamma$ and $\alpha$, the minimum and maximum degree $k_{\textrm{min}}$ and $k_{\textrm{max}}$. We use six instances of the model, with parameters chosen as $N = \{1000; 10,000\}$, $\gamma = \{2.1; 2.5; 2.9\}$, $\alpha = 2$, $k_{\textrm{max}} = 100$.
    We use either $k_{\textrm{min}} = 2$ or $k_{\textrm{min}} =3$ for the various nodes in the network to ensure an average degree $\langle k \rangle \simeq 5$.
    
\end{itemize}

\bibliography{main}

\end{document}